# Linguistic Uncertainty and Engagement in Arabic-Language X (formerly Twitter) Discourse

**Mohamed Soufan**
Independent Researcher & Data Scientist
Antalya, Turkey
ORCID: https://orcid.org/0009-0004-1705-5574
Email: m@soufan.tech

## Abstract

Linguistic uncertainty is a common feature of social media discourse, yet its relationship with user engagement remains underexplored, particularly in non-English contexts. Using a dataset of 16,695 Arabic-language tweets about Lebanon posted over a 35-day period, we examine whether tweets expressing linguistic uncertainty receive different levels and forms of engagement compared to certainty-marked tweets. We develop a lexicon-based, context-sensitive classifier to identify uncertainty markers and classify 29.9% of tweets as uncertain. Descriptive analyses indicate that uncertain tweets exhibit 51.5% higher mean total engagement (likes, retweets, and replies). Regression models controlling for tweet length, URL presence, and account verification status confirm a positive association between uncertainty and engagement ($β = 0.221$, $SE = 0.044$, $p < 0.001$), corresponding to approximately 25% higher expected engagement. The association is strongest for replies, followed by retweets and likes, suggesting a shift toward more conversational forms of engagement. Results are robust to alternative model specifications and adjustments for within-account correlation. These findings suggest that linguistic uncertainty may function as an interactional cue that encourages participatory engagement in Arabic-language social media discourse. The study contributes computational approaches for modeling linguistic features in large-scale, non-English digital communication.

## Introduction

Social media platforms have become central spaces for public discourse, where linguistic choices shape how information circulates and how audiences respond. Prior research has extensively examined the role of content features such as sentiment, emotional intensity, controversy, and misinformation in driving online engagement, particularly on platforms such as X (formerly Twitter) [1,2]. However, less attention has been paid to how linguistic uncertainty—expressions of doubt, incompleteness, or lack of verified information—affects not only the volume of engagement, but the form that engagement takes.

Linguistic uncertainty is a common feature of online communication, particularly in contexts characterized by rapid information flows, political instability, or unfolding events. Expressions such as questions, hedges, and rumors may signal informational gaps rather than firm claims, potentially inviting audience responses. While uncertainty is often treated as a weakness in information quality or as a marker of rumor diffusion, it may also function as an interactional cue that encourages users to contribute, speculate, or seek clarification [3].

Existing studies on social media engagement largely conceptualize engagement as a unified outcome, typically measured through aggregated counts of likes, shares, or comments. This approach obscures important differences between passive forms of engagement (e.g., likes) and active, conversational forms (e.g., replies). Prior work has shown that replies and retweets reflect distinct interactional behaviors, with replies more closely associated with conversational exchange [4]. As a result, we know relatively little about how linguistic cues shape not just the amount of engagement, but the interactional structure of engagement on social media.

In this study, we examine how linguistic uncertainty influences engagement patterns in Arabic-language discourse on X, focusing on tweets related to Lebanon over a 35-day period. Using a large-scale observational dataset of 16,695 tweets and a context-sensitive uncertainty classifier, we compare engagement levels between uncertain and certain tweets and assess whether uncertainty remains associated with higher engagement after controlling for key tweet-level features.

Our analysis reveals that tweets expressing linguistic uncertainty are associated with significantly higher engagement than certainty-marked tweets, with the strongest association observed for replies. We refer to this pattern as the Uncertainty–Reply Asymmetry, whereby uncertainty-marked messages are linked to disproportionately higher levels of conversational engagement relative to other engagement forms. These findings suggest that linguistic uncertainty operates not merely as an informational deficit, but as an interactional cue that may help explain how audiences engage with content in digital public spheres, consistent with interactional accounts of epistemic stance [5].

By examining how linguistic uncertainty is associated with engagement patterns, this study makes two primary contributions. First, it introduces a computationally operationalizable framework for detecting linguistic uncertainty in Arabic-language social media using an interpretable lexicon-based approach validated against human annotation. Second, it demonstrates how linguistically grounded features can be operationalized and incorporated into quantitative models to explain variation in engagement dynamics at scale. From a computational social science and information science perspective, the study shows how theory-informed language features can function as measurable signals within digital information systems, particularly in non-English contexts that remain underrepresented in large-scale modeling research.

Related Work and Theoretical Framework

Research on social media engagement has identified a wide range of content and contextual factors associated with audience response, including emotional intensity, sentiment polarity, moral language, controversy, and misinformation. Across platforms, engagement is commonly operationalized as aggregated behavioral signals such as likes, shares, and replies, which are often treated as interchangeable indicators of attention or popularity. While this literature has yielded important insights into what drives engagement volume, it has paid comparatively less attention to differences between types of engagement and the interactional dynamics they represent.

A growing body of work distinguishes between passive and active forms of engagement. Likes and retweets primarily function as low-cost signals of approval or endorsement, whereas replies require greater cognitive and communicative effort and reflect conversational involvement. This distinction suggests that engagement is not a unitary outcome, but a heterogeneous set of behaviors with different social meanings. However, few studies have examined how specific linguistic features systematically shape the composition of engagement, particularly in non-English and regionally grounded contexts.

Linguistic uncertainty has been studied across disciplines, including linguistics, psychology, and communication, where it is often associated with hedging, epistemic modality, and expressions of incomplete or provisional knowledge. In online environments, uncertainty is frequently discussed in relation to misinformation, rumor diffusion, or credibility assessment. From this perspective, uncertainty is typically framed as a deficit—either in information quality or speaker commitment. Yet this view overlooks the potential interactional functions of uncertainty in conversational settings.

From an interactional standpoint, expressions of uncertainty may serve as invitations for participation. Questions, hedges, and rumor markers can signal openness, incompleteness, or a lack of authoritative closure, thereby lowering barriers for audience response. Rather than positioning the speaker as a final authority, uncertainty-marked messages may implicitly solicit clarification, confirmation, or speculation from others, reflecting broader accounts of epistemic stance and interactional meaning [6]. This suggests that uncertainty may shape not only whether users engage, but how they engage.

Building on this perspective, we conceptualize linguistic uncertainty as an interactional cue that reconfigures engagement patterns from broadcast-oriented responses toward conversational exchange. We propose that uncertainty-marked messages are more likely to elicit replies than certainty-marked messages, even when controlling for other content and account-level features. We refer to this pattern as the *Uncertainty–Reply Asymmetry*, emphasizing the disproportionate increase in reply-based engagement relative to likes and retweets.

By foregrounding the interactional consequences of linguistic uncertainty, this study extends prior work on social media engagement in two ways. First, it shifts analytical attention from engagement volume to engagement composition. Second, it highlights the role of linguistic form in shaping participatory dynamics within digital public discourse, particularly in Arabic-language contexts that remain underrepresented in computational social science research.

## Materials and Methods

Data Collection

Data were collected from X using the Apify platform, an automated web-scraping tool that interfaces with the X web interface for data retrieval. Only publicly accessible tweets were collected in accordance with platform terms of service. Tweets were retrieved using the query (بيروت OR لبنان) with the language filter set to Arabic (lang:ar). The collection period spanned 35 consecutive days, from 15 December 2025 to 18 January 2026.

This period coincided with heightened political and economic discussion in Lebanon, making it a context in which informational uncertainty and public speculation were particularly salient. This temporal focus provides a theoretically relevant setting for examining how linguistic uncertainty operates in dynamic information environments.

The initial dataset consisted of 17,343 tweets. During preprocessing, duplicate tweets were identified and removed based on tweet identifiers, resulting in the exclusion of 648 duplicate entries. No pure retweets were present in the dataset after applying retweet filtering criteria. Quote tweets were retained, although none were present in the final dataset. Replies were included and constituted a substantial portion of the data (6,872 tweets, 41.2%). The final analytical sample comprised 16,695 tweets.

Engagement Measures

Engagement was operationalized as the sum of likes, retweets, and replies for each tweet. This composite measure reflects overall audience response while allowing for the examination of individual engagement components in descriptive analyses. Due to the highly skewed distribution of engagement values, the dependent variable in regression analyses was specified as the natural logarithm of one plus total engagement, *log(1 + Total Engagement)*.

Linguistic Uncertainty Classification

Linguistic uncertainty was identified using a rule-based uncertainty dictionary (lexicon) designed for Arabic-language content and implemented as a context-sensitive classifier. The classifier consisted of 60 uncertainty markers grouped into six categories, including modal expressions, hedging terms, question markers, contextual "who"-questions, information uncertainty phrases, and rumor indicators. The classifier incorporated context-sensitive rules to reduce false positives, such as counting the particle "من" only in interrogative contexts and excluding prepositional uses (e.g., "من بيروت").

The uncertainty lexicon comprises markers drawn from several linguistically motivated categories reflecting different forms of epistemic and informational uncertainty. Table 1 provides representative examples.

Table 1. Categories of linguistic uncertainty markers used in the classifier

| Category (Number of markers) | Examples (Arabic) | Examples (English) | Linguistic Functions |
|---|---|---|---|
| Modal verbs (6) | قد، ربما، ممكن، يمكن | may, perhaps, possible, can | Epistemic possibility |
| Hedges (12) | أعتقد، يبدو، أظن، على ما يبدو | I think, seems, I suppose, apparently | Speaker uncertainty |
| Questions (7) | هل، ماذا، لماذا، كيف، ؟ | what, why, how, ? | Information seeking |
| Contextual Who-questions (18) | من؟، من هو، مين المسؤول | who?, who is he, who is responsible | Agent/source identification |
| Information uncertainty (9) | غير متأكد، ما بعرف، مش واضح | not sure, I don't know, unclear | Explicit doubt |

| Category (Number of markers) | Examples (Arabic) | Examples (English) | Linguistic Functions |
|---|---|---|---|
| Rumor markers (8) | إشاعة، يقال، مصادر، غير مؤكد | rumor, it is said, sources, unconfirmed | Unverified information |

The uncertainty lexicon and classification rules used in this study are provided as supplementary materials.

Tweets were classified as uncertain if they contained one or more uncertainty markers. Classification performance was subsequently evaluated against human annotation (see Validation of the Uncertainty Classifier). Using this approach, 4,997 tweets (29.9%) were classified as expressing linguistic uncertainty, while 11,698 tweets (70.1%) were classified as certain.

The uncertainty dictionary and classification rules were fixed prior to analysis and remained unchanged throughout the study.

Validation of the Uncertainty Classifier

To assess the performance of the lexicon-based uncertainty classifier, we conducted a validation study using a stratified random sample of 200 tweets (100 classified as uncertain and 100 as certain). A native speaker of Lebanese Arabic annotated all tweets for the presence of linguistic uncertainty.

The classifier achieved an overall accuracy of 73.5%, with perfect recall (1.00) and moderate precision (0.47), yielding an F1-score of 0.639 and moderate agreement beyond chance (Cohen's κ = 0.470). These results indicate that the classifier successfully identified all tweets containing human-recognized uncertainty but tended to over-predict uncertainty in some contexts, particularly in narrative-style news updates and elliptical reporting structures.

Importantly, this pattern of errors implies that misclassification primarily introduces false positives (certain tweets labeled as uncertain) rather than false negatives. As a result, any measurement error is likely to attenuate observed differences between uncertain and certain tweets, meaning the observed association is unlikely to be artificially inflated by classification error and may represent a conservative estimate.

While the classifier's precision could be improved through supervised learning or deeper contextual modeling, its performance is consistent with prior lexicon-based approaches in large-scale computational social science and is sufficient for detecting systematic engagement differences at scale.

Control Variables

To account for alternative factors known to influence engagement, the regression models included the following controls: tweet length (measured as the number of characters), the presence of a URL (binary indicator), and account verification status (binary indicator). These variables were included to isolate the association between linguistic uncertainty and engagement.

Statistical Analysis

The primary analysis employed ordinary least squares (OLS) linear regression to estimate the association between linguistic uncertainty and engagement. The model was specified as:

$\log(1 + \text{Total Engagement}) = \beta_0 + \beta_1(\text{Uncertainty}) + \beta_2(\text{Tweet Length}) + \beta_3(\text{Has Link}) + \beta_4(\text{Verified}) + \varepsilon$

Because multiple tweets were posted by the same accounts, standard errors were clustered at the author level to account for potential within-account correlation (7,593 unique accounts; average 2.2 tweets per account). Statistical significance was assessed using two-tailed tests.

All analyses were conducted in Python 3.11 using the pandas, numpy, scipy, and statsmodels libraries.

As a robustness check, we also estimated negative binomial regression models predicting raw engagement counts. The negative binomial specification accounts for overdispersion in count data, a common feature of social media engagement metrics [7]. Model covariates and clustering procedures were identical to those used in the primary OLS specification.

Ethical Considerations

This study analyzed publicly available tweets only. No private accounts, protected content, or personal messages were accessed. All analyses are reported at the aggregate level, and no individual tweets, usernames, or identifying information are disclosed. The study complies with X's Terms of Service for academic research and adheres to standard ethical guidelines for the analysis of public social media data.

# Results

Descriptive Engagement Differences

We first examine descriptive differences in engagement between tweets expressing linguistic uncertainty and those expressing certainty. Of the 16,695 tweets analyzed, 4,997 (29.9%) were classified as uncertain and 11,698 (70.1%) as certain.

Uncertain tweets received substantially higher total engagement (M = 23.67) than certain tweets (M = 15.62), representing a 51.5% increase in average engagement. This pattern was consistent across individual engagement components. Compared to certain tweets, uncertain tweets received 47.0% more likes (18.84 vs. 12.82), 65.2% more retweets (2.66 vs. 1.61), and 81.5% more replies (2.16 vs. 1.19). The largest relative difference was observed for replies, suggesting a particularly strong association between linguistic uncertainty and conversational engagement.

These differences across engagement types are visualized in Figure 1.

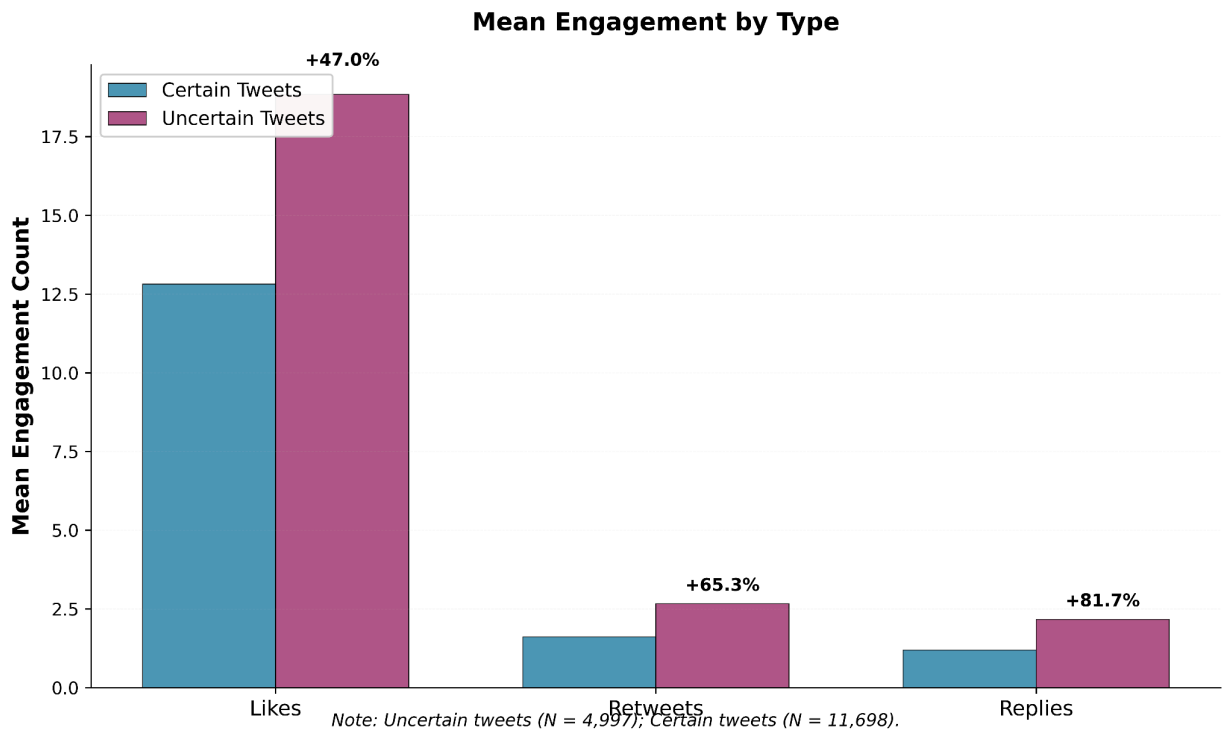

**Figure 1.** Mean likes, retweets, and replies for tweets expressing linguistic uncertainty (N = 4,997) and certainty (N = 11,698). Percent values indicate the relative increase in engagement for uncertain compared to certain tweets.

Regression Analysis

To assess whether the observed engagement differences persist after accounting for alternative explanations, we estimated an ordinary least squares (OLS) regression model predicting log-transformed total engagement. The model included linguistic uncertainty as the primary independent variable and controlled for tweet length, URL presence, and account verification status. Standard errors were clustered at the author level to account for potential non-independence arising from multiple tweets posted by the same accounts.

The results indicate a positive and statistically significant association between linguistic uncertainty and engagement. Tweets expressing uncertainty received higher engagement than certain tweets ($\beta = 0.221$, SE = 0.044, 95% CI [0.135, 0.306], $p < 0.001$). Given the log-transformed dependent variable, this coefficient corresponds to approximately a 25% increase in expected total engagement for uncertain tweets, holding other variables constant.

Among the control variables, tweet length was positively associated with engagement ($\beta = 0.001$, $p = 0.016$), as were the presence of a URL ($\beta = 0.243$, $p = 0.007$) and account verification status ($\beta = 0.376$, $p = 0.001$). The model explains a modest proportion of variance in engagement ($R^2 = 0.049$), which is consistent with prior research on social media attention dynamics characterized by highly skewed engagement distributions. Results are robust to the exclusion of high-engagement outliers (top 1% of tweets).

Negative Binomial Regression

As a robustness check, we estimated a negative binomial regression model to account for overdispersion typical of social media engagement counts [8]. Results remained substantively consistent with the OLS estimates. The incidence rate ratio (IRR) for uncertainty was 1.301 (95% CI [1.192, 1.421], $p < 0.001$), indicating that uncertainty-marked tweets are associated with approximately 30% higher expected engagement compared to certainty-marked tweets, holding other variables constant. This effect size is comparable to the OLS estimate (approximately 25% higher engagement on the original scale), confirming that the observed association is robust to model specification.

Engagement Composition and the Uncertainty–Reply Asymmetry

While linguistic uncertainty was associated with higher engagement overall, the magnitude of this association varied across engagement types. The relative increase in replies (+81.5%) was substantially larger than the corresponding increases in retweets (+65.2%) and likes (+47.0%). This disproportionate effect on replies indicates that uncertainty is more strongly associated with conversational forms of engagement than with passive or broadcast-oriented responses.

We refer to this pattern as the Uncertainty–Reply Asymmetry, reflecting the tendency of uncertainty-marked tweets to elicit higher levels of reply-based engagement relative to other engagement modalities.

## Discussion

This study examined how linguistic uncertainty is associated with engagement patterns in Arabic-language discourse on X, with particular attention to differences between engagement types. The findings indicate that tweets expressing linguistic uncertainty are associated with higher overall engagement than certainty-marked tweets, even after controlling for tweet length, URL presence, and account verification. Importantly, the association between uncertainty and engagement is not uniform across engagement modalities, but is strongest for replies, followed by retweets and likes.

The disproportionate increase in replies suggests that linguistic uncertainty may function as an interactional cue rather than merely as a content feature associated with visibility or popularity. While likes and retweets primarily function as low-cost signals of approval or dissemination, replies require active participation and conversational effort. The fact that uncertainty is most strongly associated with replies supports the interpretation that uncertainty-marked messages may invite audience involvement, prompting users to respond, clarify, or speculate rather than simply endorse or amplify content.

This pattern, which we refer to as the Uncertainty–Reply Asymmetry, highlights an important distinction between engagement volume and engagement composition. Although prior research has largely treated engagement as an aggregated outcome, the present findings demonstrate that linguistic form is associated with differences in how users engage. In this case, uncertainty is associated with a shift in engagement away from purely broadcast-oriented behaviors toward conversational interaction.

The observed difference between descriptive and regression-based estimates further underscores this point. While uncertain tweets exhibit a 51.5% higher average engagement in raw comparisons, the adjusted regression estimate indicates an approximately 25% increase after accounting for other tweet-level factors. This suggests that part of the raw engagement gap reflects correlated features such as tweet length or link presence, but that linguistic uncertainty itself remains a substantial and independent predictor of engagement. The persistence of the association after adjustment strengthens the interpretation that uncertainty is meaningfully related to interaction patterns.

These findings also contribute to broader discussions about uncertainty in online communication. Uncertainty is often framed as a liability associated with misinformation, ambiguity, or reduced credibility. However, the present results suggest that uncertainty can also serve a social function by signaling openness and incompleteness. In contexts characterized by rapid information flows and evolving events, such signals may lower the threshold for participation, encouraging users to engage in collective sense-making rather than passive consumption.

From an information science perspective, uncertainty markers can be understood as metadata-like signals embedded in natural language that influence how information objects (tweets) are interpreted, propagated, and responded to within digital information systems. By operationalizing uncertainty as a computational feature, this study demonstrates how qualitative aspects of language can be transformed into structured variables for modeling information behavior. This perspective highlights the role of linguistically encoded epistemic cues in shaping patterns of information interaction, extending analyses of information dynamics beyond network structure and topical content to include measurable properties of language itself.

Finally, the relatively modest explanatory power of the regression model ($R^2 = 0.049$) should be understood in light of the inherent unpredictability of social media engagement. Attention dynamics on platforms like X are shaped by numerous unobserved factors, including timing, network structure, and algorithmic exposure, which have been documented in prior large-scale analyses of information diffusion [9]. Within this context, identifying a consistent and interpretable association between a linguistic feature and engagement behavior represents a meaningful contribution, even if much of the overall variance remains unexplained.

## Limitations and Ethics

Limitations

Several limitations should be considered when interpreting the findings of this study. First, linguistic uncertainty was identified using a rule-based, lexicon-driven classifier. Although validation against human annotation indicates acceptable performance for large-scale analysis, the classifier demonstrated moderate precision and a tendency to over-predict uncertainty in certain contexts. As a result, some degree of misclassification remains possible. Future research could strengthen this approach through supervised learning methods or more context-sensitive modeling techniques.

Second, the analysis is observational and cannot establish causal relationships. While the regression models control for several tweet-level characteristics, unobserved factors such as posting time, network position, follower count, or algorithmic amplification may also influence engagement. As a result, the observed associations should be interpreted as correlational rather than causal.

Third, the study focuses on Arabic-language tweets related to Lebanon within a specific 35-day period. Although this context is theoretically informative, the findings may not generalize to other languages, topics, platforms, or time periods.

Finally, although standard errors were clustered at the author level to account for within-account dependence, future work could employ multilevel or mixed-effects models to more explicitly model user-level heterogeneity and capture cross-level interactions between linguistic features and user characteristics.

Ethical Considerations

This study analyzed publicly available tweets collected from X. No private accounts, protected content, or personal communications were accessed. All analyses were conducted at the aggregate level, and no individual tweets, usernames, or identifying information are reported in the paper.

The research complies with X's Terms of Service for academic research and adheres to established ethical guidelines for the use of public social media data. Given the public nature of the data and the absence of personal identification, the study poses minimal risk to users.

# Conclusions

This study examined how linguistic uncertainty is associated with engagement patterns in Arabic-language discourse on X, focusing on tweets related to Lebanon. Using a large observational dataset and a context-sensitive uncertainty classifier, the analysis shows that tweets expressing linguistic uncertainty are associated with significantly higher engagement than certainty-marked tweets. Importantly, this association is not uniform across engagement types: uncertainty is most strongly associated with replies, indicating a shift toward conversational participation rather than passive endorsement.

By distinguishing between engagement volume and engagement composition, the study shows that linguistic uncertainty functions as more than an informational deficit. Instead, uncertainty is associated with patterns consistent with its role as an interactional signal that invites response, encouraging users to participate in discussion, clarification, or collective sense-making. We conceptualize this pattern as the *Uncertainty–Reply Asymmetry*, highlighting the disproportionate role of replies in uncertainty-driven engagement.

These findings contribute to research on digital discourse and computational social science by demonstrating that linguistic form can systematically influence how audiences engage, not only how much they engage. In contrast to approaches that treat engagement as a single aggregated outcome, this study underscores the value of examining engagement modalities separately to better capture interactional dynamics.

More broadly, the results suggest that uncertainty plays a complex role in online public communication. While often associated with ambiguity or reduced credibility, uncertainty may also foster participation in contexts characterized by evolving information and contested interpretations. Understanding this dual role is particularly relevant for analyzing discourse in high-salience political and social environments.

Future research can build on these findings by validating uncertainty classifications through human annotation, examining the qualitative content of replies, and testing whether similar interactional patterns emerge across languages, topics, and platforms. Together, these extensions would further clarify how linguistic uncertainty shapes participation in digital public spheres.

## Supplementary Materials

The Arabic uncertainty lexicon and context-sensitive classification rules are provided as Supplementary Materials.

## Data Availability Statement

Due to platform policy and privacy considerations, raw tweet text cannot be shared. Tweet identifiers and derived aggregate variables used in the analyses are available from the corresponding author upon reasonable request. The uncertainty lexicon and classification rules are provided in the Supplementary Materials. The analysis code used in this study is also available from the corresponding author upon reasonable request.

## Author Contributions

Mohamed Soufan conducted all aspects of the study, including conceptualization, methodology, software development, formal analysis, investigation, data curation, visualization, and writing. Manual annotation assistance used for validation is acknowledged below.

## Acknowledgments

The author thanks Aya Soufan for her assistance with manual annotation during the validation of the linguistic uncertainty classifier.

## Funding

This research received no external funding.

Institutional Review Board Statement

Ethical review and approval were waived for this study because the research analyzed publicly available social media data and did not involve interaction with human participants or access to private or sensitive information.

Informed Consent Statement

Not applicable

Conflicts of Interest

The author declares no conflict of interest.